# Glassy dynamics in mono-, di-, and tri-propylene glycol: From the α- to the fast β-relaxation

M. Köhler[1], P. Lunkenheimer[1], Y. Goncharov[2], R. Wehn[1], A. Loidl[1]

[1] *Experimental Physics V, Center for Electronic Correlations and Magnetism, University of Augsburg, 86135 Augsburg, Germany*
[2] *Institute of General Physics, Russian Academy of Sciences, 119991 Moscow, Russia*

**Abstract**

We present a thorough characterization of the glassy dynamics of three propylene glycols (mono-, di- and trimer) by broadband dielectric spectroscopy. By covering a frequency range of more than 15 decades, we have access to the entire variety of dynamic processes typical for glassy dynamics. These results add three more molecular glass formers to the sparse list of materials for which real broadband spectra, including the region of the fast $\beta$-process, are available. Some first analyses of the various observed dynamic processes are provided.

## 1. Introduction

The dynamics of glassy matter reveals a variety of unusual and so far only poorly understood phenomena. A vast number of publications treats the structural $\alpha$-relaxation, the "excess wing" and the Johari-Goldstein (JG) relaxation [1]. While even the $\alpha$-relaxation, known since the beginning of the research of glassy dynamics, still involves many open issues, the microscopic origin of the latter two features is unclarified at all. Sometimes excess wing and JG-relaxation peak are discussed on the same footing [2,3] but also this notion is not finally settled.

Another typical feature of glassy matter is the so-called fast $\beta$-relaxation, showing up in the GHz-THz range. This region attracted much interest during recent years as the fast dynamics may be the key for a better understanding of the glass transition and the glassy state of matter in general. By dielectric spectroscopy this high frequency region is only accessible under great effort and likewise extremely challenging. For this reason, in the high frequency regime only sparse data can be found in literature [4,5,6,7,8,9,10]. The fast $\beta$-relaxation is predicted by mode coupling theory (MCT) and ascribed to the "rattling" motion of a particle in the transient cage formed by its neighbouring molecules [11]. The fast $\beta$–process usually is assumed to show up as a shallow minimum in the GHz-THz region. There are also other approaches to explain the excess intensity observed in this high-frequency region. Maybe the most prominent one is the extended coupling model [12], which involves an explanation in terms of a nearly constant loss, a phenomenon which is known since long [13]. Unfortunately, so far only two molecular and two ionic glass formers have been investigated by dielectric spectroscopy in a sufficiently broad and continuous frequency range to allow for a meaningful analysis of the fast $\beta$-relaxation [5,7,8,9,10].

Many groups studied the dynamics of propylene glycols, both the monomer as well as glycols exhibiting different grades of polymerisation. Different techniques were used for characterization, among them neutron [14] and light scattering [15,16,17,18,19,20], optical Kerr-effect [21] as well as photothermal [22] and dielectric spectroscopy [3,23,24,25,26,27,28,29,30] including elaborated studies at elevated pressure [26-29]. Moreover, different models were applied to the experimental data [31], especially the coupling model [12], the minimal model [32,33] and a nonmonotonic relaxation kinetic model [34]. However, in most dielectric experiments reported so far, the frequency range was restricted to frequencies below 10 MHz, except for [24] where spectra up to 1 GHz were reported.

In the present work we show broadband spectra ranging from mHz up to the THz region for propylene glycol (PG) and its di- and trimer (DPG and TPG, respectively). The main purpose of the present work is to provide real broadband spectra, including the fast-$\beta$ region for three more molecular glass formers, in addition to the two materials investigated so far (glycerol and propylene carbonate [5,9,10]). As the investigated glycols are closely related, one can expect information on the influence of the molecular size on glassy dynamics. In addition, a first phenomenological analysis of all the observed processes of glassy dynamics is presented and a preliminary evaluation of the fast $\beta$-relaxation in terms of the standard version of MCT is performed.

**2. Experimental procedures**

A combination of different experimental techniques is necessary to record the real and imaginary part of the dielectric permittivity in the broad frequency range covered by the present work. For details the reader is referred to [5, 35]. Propylene glycol was purchased from Fluka with a purity $\geq$ 99.5%, dipropylene glycol from Aldrich with a purity > 99% and tripropylene glycol from Alfa Aesar with a purity $\geq$ 98.91%. All samples were used without further purification. Glass transition temperatures were determined by differential scanning calorimetry (DSC) using a *Perkin-Elmer DSC Pyris 1* using cooling and heating rates of 10 K/min.

**3. Results and Discussion**

In Figure 1, the frequency-dependent dielectric loss $\varepsilon''(\nu)$ of (a) PG, (b) DPG and (c) TPG is shown for various temperatures. All spectra are dominated by the well-known $\alpha$-relaxation peaks, whose strong temperature-dependent shift displays the tremendous slowing down of glassy dynamics towards the glass transition. In addition, at frequencies beyond the $\alpha$ relaxation an excess wing is observed for PG whereas the dimer and trimer reveal well-pronounced secondary relaxation peaks [3,28,30]. With increasing temperature, the latter become successively submerged under the dominating $\alpha$ peak leading to an excess wing at about 210 K. Finally, above about 240 K, $\alpha$ and $\beta$ relaxation are merged. In the last two frequency decades, a shallow minimum shows up in all three materials. Its magnitude increases and it shifts to higher frequencies with increasing temperature as found also in other glass formers [5,7,8,9,10].

*3.1 The $\alpha$ relaxation*

To obtain information on the characteristic parameters of the $\alpha$ and $\beta$ relaxation, least square fits of the experimental data were simultaneously performed for the imaginary *and* real part (not shown) of the dielectric permittivity. For parameterization of the spectra at frequencies below the minimum region, a sum of the commonly used Cole-Davidson (CD) [36] and Cole-Cole (CC) functions [37] was used. The former characterizes the $\alpha$ relaxation and the latter the slow $\beta$-processes at temperatures below 220 K in PG and TPG and 230 K in DPG. It should be mentioned that the justification for using an additive superposition ansatz may be doubted and alternatives were promoted [38]. However, for a first parameterisation such an ansatz works well and usually the convolution approach proposed, e.g., in [38] leads to very similar results (see, e.g., [6]). It should be noted that in the present work the use of the term "$\beta$ relaxation" for the secondary relaxations in all three glycols has only practical reasons; in fact there is reason to believe that they are not due to the same microscopic mechanism (cf. section 3.2). For PG, here we adopt the picture that the excess wing is due to a $\beta$ relaxation [2,3].

|  | PG | DPG | TPG |
|---|---|---|---|
| $\tau_0$ (fs) | 14 | 25 | 53 |
| $T_{VF}$ (K) | 115 | 150 | 150 |
| D | 17 | 10 | 9 |
| *m* | 48 | 69 | 74 |
| $T_{CW}$ (K) | 53 | 105 | 95 |
| $T_g$(DSC) | 170 | 196 | 194 |
| $T_g$(diel.) | 166 | 189 | 193 |

Table 1. Relaxation parameters of the investigated glycols. In addition, the glass temperatures determined by DSC and from the condition $\tau_\alpha(T_g) = 100$ s are provided. For definition of parameters see text.

The fits yield the relaxation time $\tau_{CD}$ and the width parameter $\beta_{CD}$ of the CD function, from which the average relaxation time $<\tau_\alpha> = \tau_{CD} \beta_{CD}$ can be calculated [5]. Its temperature dependence is shown in Fig. 2 (closed symbols). The phenomenological Vogel-Fulcher-Tammann function, $<\tau_\alpha> = \tau_0 \exp[DT_{VF} / (T-T_{VF})]$, well accounts for the observed non-Arrhenius behaviour, typical for glassy dynamics. $T_{VF}$ is the Vogel-Fulcher temperature, $D$ the strength parameter [39] and $\tau_0$ a prefactor. Resulting fit parameters are given in Table 1. They all agree reasonably well with the values reported in [23,24,30].

The fragility parameter *m*, defined as the slope at $T_g$ in the Angell-plot [40] (inset of Figure 2), gives evidence for intermediate fragility, increasing with increasing molecular size from $m = 48$ to 74. León *et al.* reported similar values ($m = 53$, 64 and 71 for PG, DPG and TPG respectively) [30]. In [41] the fragility was related to the form of the potential energy landscape in configuration space. Within this framework, the observed increase in fragility of the glycols may be ascribed to a higher density of minima in the potential energy landscape for the more complex molecules.

Figure 2 reveals that, contrary to naive expectations, the relaxation of the larger trimer molecules seems to be faster than for the dimer. As an explanation one may speculate that the relaxation of the trimer already indicates a transition to a polymer-like dynamics where usually segmental motions are the dominant relaxation processes. Also weakening of the intermolecular hydrogen bonding due to extended molecular size could provide an explanation. Interestingly, scaling the relaxation times in the Angell plot (inset of Fig. 2) makes this puzzling behaviour vanish.

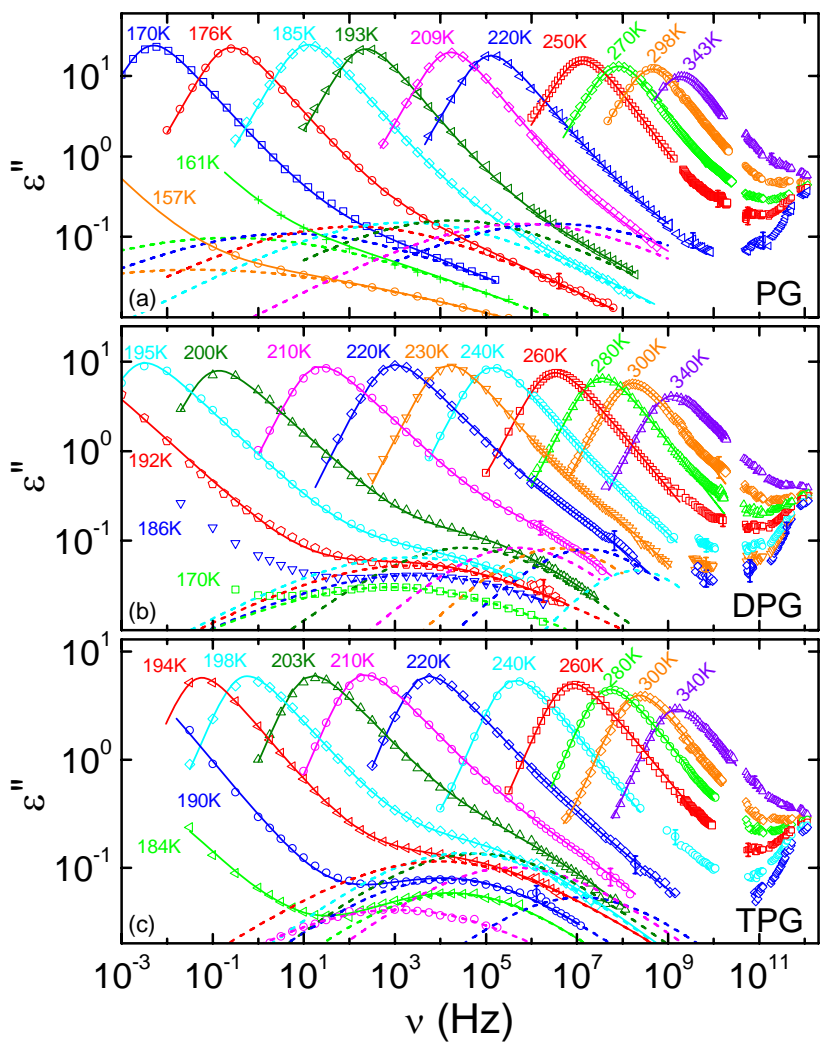

Fig. 1. Frequency-dependent dielectric loss of (a) PG, (b) DPG and (c) TPG for selected temperatures. The solid lines are fits with the sum of a CD and a CC function. (At $T \geq 220$ K for PG and TPG and $T \geq 230$ K for DPG the CC amplitude was set to zero.) The dotted lines show the CC parts of the fits. Data on PG are partly published in Ref. 3. The errors are of similar magnitude as the symbol size, except where explicitly indicated.

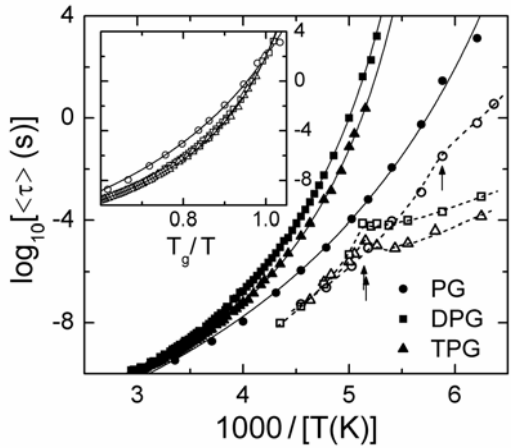

Fig. 2. Relaxation times determined from the fits shown in Fig. 1. The solid lines are fits of the α-relaxation times (closed symbols) with a VFT function. The open symbols correspond to the β-relaxation times. The dashed lines are guides to the eye. The arrows indicate $T_g$. The inset shows an Angell-plot of the α-relaxation times.

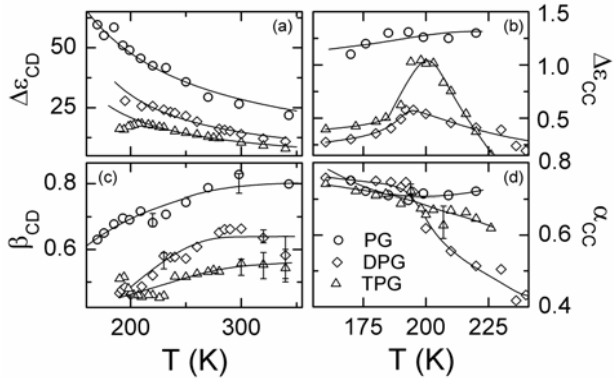

Fig. 3. Relaxation strengths of the α (a) and β (b) process as obtained from the fits of the dielectric loss spectra shown in Fig. 1. Solid lines in (a) are fits using the Curie-Weiss law. (c) and (d) show the width parameters for α (c) and β (d) relaxation. Lines in (b), (c) and (d) are drawn to guide the eyes.

The relaxation strength $\Delta\varepsilon_{CD}$ and width parameter $\beta_{CD}$ of the α relaxation are shown in Figs. 3(a) and (c), respectively. At high temperatures a saturation of $\beta_{CD}$ below unity is observed for all three glycols. Such behavior was also found in other molecular glass-formers [5,6]. $\Delta\varepsilon_{CD}$ increases with decreasing temperature and molecular size (Fig. 3(a)). Its temperature dependence can be best parameterised by a Curie-Weiss law, $\Delta\varepsilon \sim 1/(T-T_{CW})$ (lines). For DPG and TPG some deviations show up when approaching 200 K. This can be ascribed to the fact that the glass-transition temperatures for both compounds are in this region (cf. Tab. 1) and, thus, the sample falls out of thermodynamic equilibrium. A temperature dependence of the relaxation strength stronger than a simple Curie law ($\Delta\varepsilon \sim 1/T$) is often observed in glass-forming matter [9,42] and may point to increasingly cooperative relaxation for decreasing temperature.

### 3.2 The secondary relaxation

The secondary relaxations of the glycols were extensively investigated in various earlier publications (e.g., [3,24,27]) and, thus, here we provide only a brief discussion. In [3], the secondary relaxation of PG was classified as a genuine JG relaxation [1], whereas the secondary peaks in DPG and TPG where identified as non-JG [24,27]. JG β-relaxations are assumed to be inherent to the glassy state and seem to show some properties that are common to all glass formers. In contrast, non-JG secondary relaxations are regarded as material-specific, often ascribable to intramolecular modes, and of no deeper interest for the understanding of the glass transition. In the di- and trimer, high pressure experiments revealed an additional third relaxation process [27]. It is located in the frequency range between the α and β-process and was identified as the JG process [27]. Thus the β-processes observed in the present work seem to be of JG type for PG and of intramolecular type in DPG and TPG.

The β-relaxation times obtained from the fits in Fig. 1 are shown in Fig. 2 (open symbols). For many glass formers, in non-equilibrium at $T < T_g$ an Arrhenius behaviour of the β–relaxation times is well established. However, for $T > T_g$, data analysis is hampered by strong merging of both processes and, thus, the behavior in this region is not finally clarified [43]. We show here the results of our additive fits, which we believe to at least provide a rough estimate of the behavior at $T > T_g$. Above $T_g$ (indicated by the arrows), $\tau_\beta(T)$ shows a transition into a stronger temperature dependence than at $T < T_g$. Similar behavior was also found in various other glass formers [3,6,44]. It is interesting that the β dynamics for all three glycols approaches each other with increasing temperature. This seems to contradict a different nature of the secondary relaxations in PG and in DPG/TPG [24,27] but, clearly, further investigations are necessary to clarify this issue. In the sub-$T_g$ region, a minimum in $\tau_\beta(T)$ is observed for DPG and TPG (Fig. 3). Grzybowska et al. [28, 33] described this minimum in the framework of the minimal model [32] and a nonmonotonic relaxation kinetic model [34].

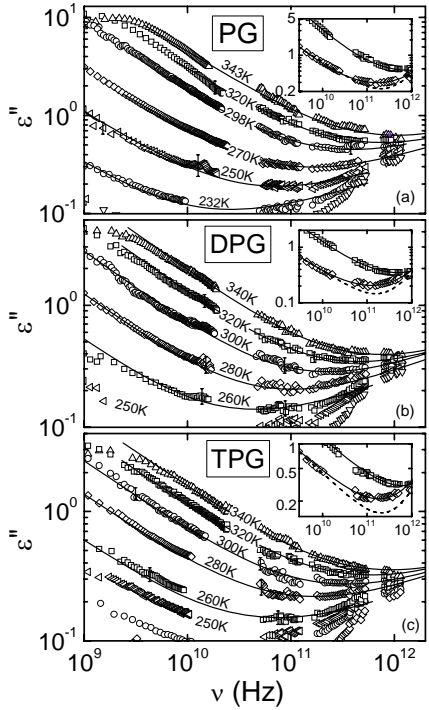

Fig. 4. $\varepsilon''(\nu)$ of PG (a), DPG (b) and TPG (c) at high frequencies for various temperatures. The lines are fits with MCT, eqs. (2) and (3). The insets show spectra for two temperatures, fitted by eq. (1) (solid lines). In addition, curves calculated by the sum of two power laws are shown (dashed lines, see text).

The width parameters $\alpha_{CC}$ and the relaxation strengths $\Delta\varepsilon_{CC}$ of the $\beta$-relaxations are provided in Figs. 3(b) and (d). In all three materials $\alpha_{CC}(T)$ tends to decrease, i.e. the $\beta$-peaks narrow with increasing temperature, a behavior typical for secondary relaxations (see, e.g., [6,45]). The di- and trimer both show a peak in $\Delta\varepsilon_{CC}(T)$, located close to $T_g$ (Fig. 3(b)). An increase of $\Delta\varepsilon$ below $T_g$ is predicted by the minimal model [32]. It is not clear if the decrease above $T_g$ is an artefact of merging.

### 3.3. The fast $\beta$ relaxation

Glassy dynamics in the high-frequency region above GHz has found much interest during recent years. In the glycols (Fig. 4) and other glass formers [5,7,8,10], here a characteristic shallow loss minimum occurs. The interest in this region was mainly triggered by the MCT, whose main outcome is the prediction of non-trivial additional contributions to the susceptibility in the GHz-THz region. For a model-free check of such an excess intensity, the insets of Fig. 4 show curves which were calculated by the sum of a two power laws (dashed lines). The high-frequency power law was assumed to have an exponent of at least one as the increase towards the boson peak, which will appear beyond the investigated frequency region [5], should be steeper than linear [10,46]. Clearly, this ansatz is not sufficient for the description of the data and, thus, there is evidence for excess intensity in the minimum region.

One possibility for the description of this excess contribution is provided by the assumption of a constant loss $\varepsilon_c$. Using an additive superposition of the above-mentioned two power laws with exponents $b < 1$ and $n \geq 1$, one arrives at:

$$\varepsilon'' = c_b \nu^{-b} + \varepsilon_c + c_n \nu^n \qquad (1)$$

As also found in other glass formers [7,8,47], for each glycol the minimum region of $\varepsilon''(\nu)$ can be reasonably well described by eq. (1). This is shown for two temperatures by the fit curves included in the insets of Fig. 4 (solid lines). The presence of a constant loss in glassy matter has been considered since long (e.g., [13]) and quite recently found renewed interest due to its incorporation into the extended coupling model [12].

In the following, we will provide a preliminary analysis of our data in the framework of the idealized MCT; a treatment with the extended F12 model of MCT [48] will be published in a forthcoming paper. Within idealized MCT, for $T$ above but near the critical temperature $T_c$, the minimum is approximated by the sum of the von Schweidler law, $\nu^{-b}$, and the critical law, $\nu^a$ [11]:

$$\varepsilon'' = \frac{\varepsilon_{min}}{a+b}\left[ a\left(\frac{\nu}{\nu_{min}}\right)^{-b} + b\left(\frac{\nu}{\nu_{min}}\right)^{a}\right] \qquad (2)$$

Here $\nu_{min}$ and $\varepsilon_{min}$ denote position and amplitude of the minimum. The temperature-independent exponents $a$ and $b$ are constrained by the system parameter

$$\lambda = \frac{\Gamma^2(1-a)}{\Gamma(1-2a)} = \frac{\Gamma^2(1+b)}{\Gamma(1+2b)} \qquad (3)$$

where $\Gamma$ denotes the Gamma function. The fits with Eq. (2) shown in Fig. 4 provide a good description of the experimental data over 2-3 frequency decades. The agreement is at least as good as in propylene carbonate or glycerol [5,10]. At low temperatures, especially in PG some deviations of data and fits occur in the region of the critical law, similar to the findings in glycerol [5]. In the latter compound they were ascribed to the additional increase towards the boson peak, not covered by MCT.

The critical temperature should manifest itself in the temperature dependence of the $\varepsilon''(\nu)$-minimum and the $\alpha$-peak, according to the three critical laws of MCT:

$$\nu_\alpha \propto (T - T_c)^\gamma \qquad (4)$$

$$\nu_{\min} \propto (T - T_c)^{\frac{1}{2a}} \qquad (5)$$

$$\varepsilon_{\min} \propto (T - T_c)^{1/2} \qquad (6)$$

Here $\nu_\alpha$ is the relaxation rate $1/(2\pi\tau_\alpha)$. $\gamma$ is defined as $\gamma = 1/(2a) + 1/(2b)$. $\varepsilon_{\min}$ and $\nu_{\min}$, taken from the fits with eq. (2), as well as the $\alpha$-relaxation rate $\nu_\alpha$ are shown in Fig. 5 in a representation that should lead to straight lines extrapolating to $T_c$. For all three glycols the complete set of parameters can be consistently described with $T_c \approx 239$ K for PG, $T_c \approx 259$ K for DPG and $T_c \approx 250$ K for TPG (solid lines). When considering these results, one should be aware that the relations (4) – (6) should only hold above but near $T_c$ and are expected to fail too far above $T_c$. Depending on the exact choice of the extrapolation curve, different critical temperatures can be obtained. For this reason the significance of the results of Fig. 4 should not be overemphasized, although similar evaluations are often used in literature. The parameters resulting from this preliminary analysis are summarized in Tab. 2. In literature, only for the monomer some inconsistent data were reported with critical temperatures of 198, 314 and 251 K [20,42]. The ratio $T_c/T_g$ decreases from 1.41 to 1.28 with increasing molecular size. Hence it is slightly above 1.2, a value often mentioned in literature for this ratio.

|   | PG | DPG | TPG |
|---|---|---|---|
| $a$ | 0.33 | 0.319 | 0.302 |
| $b$ | 0.653 | 0.612 | 0.55 |
| $\lambda$ | 0.692 | 0.717 | 0.7556 |
| $\gamma$ | 2.28 | 2.38 | 2.566 |
| $T_c$ (K) | 239 | 259 | 250 |
| $T_c/T_g$ | 1.41 | 1.32 | 1.28 |

Table 2: Parameters as derived from MCT analysis. For details see text.

Finally it should be noted that in [49] it was proposed to scale the loss data with the static permittivity of the $\alpha$-process before applying MCT. This leads to a change of the temperature dependence of $\varepsilon_{\min}$ (open symbols in Figs. 5(a) - (c)). Also these scaled data are consistent with the values of $T_c$ given in Tab. 2

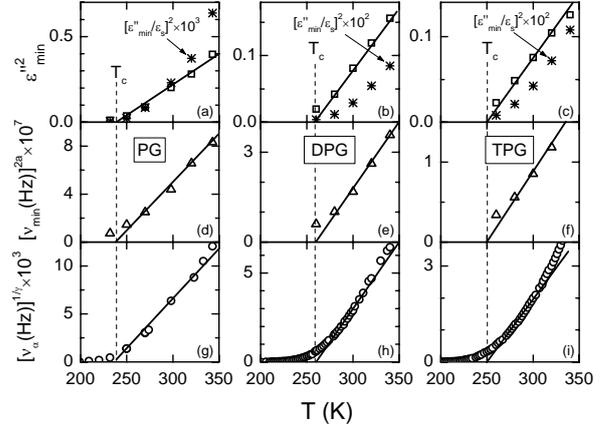

Fig. 5: Critical temperature dependence of the amplitude (a-c) and position (d-f) of the $\varepsilon''(\nu)$ minimum and of the $\alpha$-relaxation rate (g-i). The lines correspond to the critical laws of MCT, eqs. (4) - (6). In (a-c), additional results obtained from scaled loss data, as proposed in [49], are included (stars).

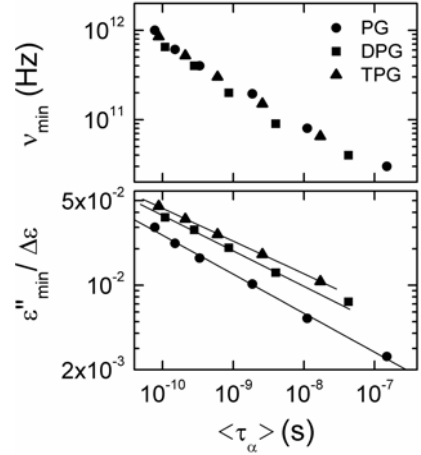

Fig. 6. Minimum position of PG, DPG and TPG (a) and minimum amplitude, relative to the $\alpha$-relaxation strength, (b) vs. $\alpha$-relaxation time. The lines in (b) are linear fits.

To compare the behavior of the fast $\beta$-relaxation of the different glycols, Fig. 6 shows the minimum frequency and the minimum amplitude, related to the $\alpha$-relaxation strength, vs. the mean $\alpha$-relaxation time $\langle\tau_\alpha\rangle$. This representation allows for a comparison of the minimum parameters at identical $\alpha$-relaxation dynamics. The $\nu_{\min}$ values of all glycols approximately fall on a single line (Fig. 6(a)). However, the relative minimum amplitude and thus the strength of the fast $\beta$-relaxation increases with increasing molecule size. This may be ascribed to a better coupling of the dielectric loss to translational modes for the larger molecules. It is well known that the relative amplitude of the fast process is weaker in dielectric spectroscopy, compared to the susceptibilities determined by scattering methods, which more directly couple to density fluctuations [5,7,9,10]. For the spatially more extended DPG and TPG molecules, steric hindrance may rise and thus reorientational motions may be more strongly coupled to translational ones.

**4. Summary and Conclusion**

In the present work, we have provided broadband dielectric spectra covering 15 decades in frequency for three glass-forming propylene glycols, thereby considerably increasing the number of glass formers for which data extending into the region of the fast $\beta$-relaxation are available. Besides the $\alpha$ relaxation, these glycols show all the characteristic high-frequency processes of glassy dynamics that have attracted so much interest in recent years. Of special significance is the finding of clear evidence for a fast $\beta$-relaxation also in these materials. It exhibits characteristics, very similar to those in glycerol and propylene carbonate, the only molecular glass formers that so far have been investigated in a similarly broad and continuous frequency range. We have performed a thorough phenomenological characterization of the $\alpha$ and slow $\beta$–relaxation of these glass-forming glycols. In addition, it was demonstrated that the excess intensity in the region of the loss-minimum can be described by a constant loss contribution. Finally, idealised MCT was shown to provide a reasonable description of the fast $\beta$-process. From this preliminary analysis, first estimates for the critical temperature and system parameters were obtained. One interesting aspect of this analysis is the increasing strength of the fast $\beta$-process with increasing molecular size, which may indicate an increasing coupling of reorientational and translational motion for the larger molecules. An analysis in the framework of the more sophisticated F12-model of MCT [48] will be provided in a forthcoming work.